\documentclass{ifacconf}

\usepackage{graphicx}      
\usepackage{natbib}        

\usepackage{amsmath}
\usepackage{amssymb}
\usepackage{bbm}

\newcommand{\tr}{\operatorname{tr}}
\newcommand{\diag}{\operatorname{diag}}
\newcommand{\st}{\, | \,}
\newcommand{\be}{\begin{equation}}
\newcommand{\ee}{\end{equation}}
\newcommand{\R}{\mathbb R}

\DeclareMathOperator{\vol}{volume}

\newtheorem{Theorem}{Theorem}
\newtheorem{Definition}{Definition}
\newtheorem{Remark} {Remark}
\newtheorem{Example} {Example}

\begin{document}
\begin{frontmatter}

\title{A sufficient condition for $k$-contraction in Lurie~systems}

\thanks[footnoteinfo]{This research was partly supported by a research grant from the~Israel Science Foundation. The work of AO was partly supported by a research grant from the Ministry of Aliyah and Integration.}

\author[First]{Ron Ofir} 
\author[Second]{Alexander Ovseevich}
\author[Third]{Michael Margaliot}

\address[First]{The Andrew and Erna Viterbi Faculty of Electrical and Computer Engineering, Technion---Israel Institute of Technology, Haifa 3200003, Israel (e-mail: rono@campus.technion.ac.il).}
\address[Second]{School of Electrical Engineering, Tel Aviv University, Israel 69978 (e-mail: ovseev@gmail.com)}
\address[Third]{School of Electrical Engineering, Tel Aviv University, Israel 69978 (e-mail: michaelm@tauex.tau.ac.il)}

\begin{abstract}                
We consider a Lurie system obtained via a connection of a linear time-invariant system and a 
nonlinear feedback function.  Such systems often have more than a single equilibrium and are thus not contractive with respect to any norm. We derive a new  sufficient condition for $k$-contraction of a Lurie system. For~$k=1$, our sufficient 
condition reduces to the standard stability condition based on the bounded real lemma and a small gain condition.
 For~$k=2$, our condition
 guarantees   well-ordered  asymptotic behaviour of the closed-loop system:   every bounded solution converges to an equilibrium, which is not necessarily unique.
 We apply   our results to derive 
 a sufficient condition for $k$-contractivity
 of a networked system. 
\end{abstract}

\begin{keyword}
Stability of nonlinear systems, contraction theory, bounded real lemma, $k$th compound matrices, Hopfield network. 
\end{keyword}

\end{frontmatter}

\section{Introduction}
Consider a nonlinear system obtained by connecting
a linear time-invariant~(LTI) system with state vector~$x\in \R^n$, input~$u\in\R^m$ and output~$y\in \R^q$:   
\begin{equation}\label{initial}
\begin{array}{l}
\dot x=Ax+ Bu ,\\
y=Cx+Du,
\end{array}
\end{equation}
with a time-varying 
nonlinear   feedback control 
$u=-\Phi(t,y).$
The resulting closed-loop system is known as the Lurie system after the Russian mathematician A. I. Lurie. Such systems play an important role in systems and control theory, as many nonlinear real-world systems
can be represented as a Lurie system. 
A non-trivial and  well-studied  topic
is proving that the closed-loop system is asymptotically stable for $\Phi$ belonging to certain   classes  of nonlinear functions, e.g., the class of sector-bounded functions~\citep[Ch.~7]{khalil_book}.  

 When~$D=0$, the closed-loop system is 
\begin{equation}\label{eq:closed_loop1}
    \dot x = Ax - B\Phi(t,Cx).
\end{equation}
In the 1940s and 1950s,    Aizerman and     Kalman
conjectured that for certain classes of nonlinear functions the stability analysis of the closed-loop system~\eqref{eq:closed_loop1} can be reduced to the stability analysis of certain classes of linear systems.  
These  conjectures are now known to be false.

Several authors studied~\eqref{eq:closed_loop1} using contraction theory.
 \cite{Smith1986haus} reformulated   a  bound on the Hausdorff  dimension of  compact invariant sets of mappings~~\citep{Douady1980}
 to dynamical systems, applied it to a Lurie system, and demonstrated the results by  bounding  the Hausdorff dimension of attractors of the Lorentz equation. However, his  results are
  highly conservative, especially for large-scale systems. \cite{Andrieu2019LMIContract}  provide a Linear Matrix Inequality (LMI) sufficient condition for contraction  w.r.t.  Euclidean norms under differential sector bound or monotonicity assumptions on the nonlinearity (see also~\cite[Theorem 3.24]{bullo_contraction} for a similar condition under different  assumptions), and use it to design controllers which guarantee contraction of  the closed-loop system. This design method was then revisited by~\cite{control_syn_vincent} where it was shown that the designed controllers yield a closed-loop system with infinite gain margin.  \cite{Proskurnikov2022GeneralizedSLemma} provide  a sufficient condition for contraction w.r.t. non-Euclidean norms (see also~\cite{AD-AVP-FB:21k} where this question was studied in the context of recurrent neural networks). However, 
a Lurie system may have more than a single equilibrium point~\citep{MIRANDAVILLATORO201876}, and then it is clearly not contractive w.r.t. any norm.

Following the seminal work   of~\cite{muldowney1990compound},
\cite{kordercont} recently
introduced  the notion of  $k$-contractive systems. 
Classical  contractivity implies  that under the phase flow of the system  the tangent vectors to the phase space contract exponentially fast.  $k$-contactivity implies that  the same property holds for   elements of $k$th exterior  powers of the tangent spaces. Roughly speaking, this is equivalent to the fact that the flow of the  variational  equation  contracts $k$-dimensional parallelotopes  at an exponential rate. In particular, a~$1$-contractive system is just a contractive system. 
However, a   $k$-contractive system, with~$k>1$, may not be contractive in the standard sense. For example, every bounded solution of a time-invariant $2$-contractive system  
converges to an equilibrium point, which may not be unique~\citep{li1995}. Thus, 2-contraction may be useful for analyzing multi-stable systems that cannot be analyzed using standard contraction theory.

The basic tools required to define and study~$k$-contractivity are the~$k$th multiplicative and~$k$th additive compounds of a  matrix. 
  The reason for this is simple: 
$k$th multiplicative compounds  provide information on the volume of  parallelotopes generated by~$k$ vertices, and 
$k$th additive compounds  describe the dynamics of $k$th multiplicative compounds  when the vertices follow
a linear dynamics.

Here, we derive and prove a novel sufficient condition for~$k$-contractivity of a Lurie system. A unique feature of this condition is that it
combines an algebraic Riccati inequality~(ARI)
that includes the $k$th additive compounds of the matrices of the~LTI, and a kind of gain condition on the Jacobian of the nonlinear function~$\Phi$, and is therefore   less conservative than the sufficient condition in~\citep{Smith1986haus}. 

In the special case~$k=1$, our condition  reduces to a small-gain sufficient condition for standard contraction. 
However, for~$k>1$ our results may be used to analyze systems that are not  contractive. 
We demonstrate  this  by deriving a sufficient  condition for~$k$-contractivity of a Hopfield network. These networks are often used as associative memories with every stored pattern corresponding  to an equilibrium of the  dynamics. Thus, they are typically multi-stable,  and cannot be analyzed using standard contraction theory.

We use     standard notation. For a square matrix~$A$, $A^T$ is the transpose of~$A$,  and $\tr (A)$ [$\det (A)$]
is the trace [determinant] of~$A$. 
A symmetric matrix~$P \in \R^{n \times n}$ is called positive definite [positive semi-definite] if $x^TPx > 0$ [$x^TPx \ge 0$] for all~$x \in \R^n\setminus\{0\}$. Such matrices are denoted by $P \succ 0$ and $P \succeq 0$, respectively. Given a matrix~$A \in \R^{n \times m}$, we use~$\sigma_1(A) \ge \dots \ge \sigma_{\min\{n,m\}}(A) \ge 0$ to 
denote the ordered singular values of~$A$. The~$n\times n$ identity matrix is denoted by~$I_n$. 
The~$L_2$  norm  of a vector~$x$ is~$|x|_2:=(x^Tx)^{1/2}$.  
For two integers~$i,j$, with~$ i\leq j$, let~$[i,j]:=\{i,i+1,\dots,j\}$.

  The next section reviews known definitions and results that  are used later on. Section~\ref{sec:main}
includes  our main results.

\section{Preliminaries}
We review several definitions and results on  matrix  compounds and matrix measures (also known as logarithmic norms~\citep{strom1975logarithmic}) that will be used in Section~\ref{sec:main}.
 \subsection{Matrix compounds}
 Let~$Q_{k,n}$ denote the set 
of increasing sequences of~$k$ numbers from~$[1,n]$
ordered lexicographically.
For example,~$Q_{2,3} =
\{ (1,2), (1,3), (2,3) \}
$. 
For~$A\in\R^{n\times m}$ and~$k\in[1,\min\{n,m\}]$, 
  a \emph{minor of order~$k$} of~$A$ is the determinant of some~$k \times k$
submatrix of~$A$.  
Consider the~$\binom{n}{k}\times \binom{m}{k}  $
 minors of  order~$k$ of~$A$. 
Each such minor is defined by a set of row indices~$\kappa^i \in Q_{k,n}$ and column indices~$\kappa^j\in 
Q_{k,m}$. This minor 
is denoted by~$A(\kappa^i|\kappa^j)$.
For example, for~$A=\begin{bmatrix} 4&5   \\ -1 &4   \\0&3 
\end{bmatrix}$,  we have
$
A(\{ 1,3\} |\{1,2\})=\det  \begin{bmatrix} 4&5\\0&3
\end{bmatrix}   =12.
$
\begin{Definition}\label{def:multi} 
    The~$k$th \emph{multiplicative compound matrix} 
of~$A\in\R^{n\times m}$, denoted~$A^{(k)}$, is the~$\binom{n}{k}\times  \binom{m}{k}$ matrix
that includes all  the minors of order~$k$ ordered lexicographically.
\end{Definition} 
For example, for~$n = m =3$ and~$k=2$, we have 
\[
	 A^{(2)}= \begin{bmatrix}
                   A(\{1,2\}|\{1,2\}) & A(\{1,2\}|\{1,3\}) & A(\{1,2\}|\{2,3\})\\
						A(\{1,3\}|\{1,2\}) & A(\{1,3\}|\{1,3\}) & A(\{1,3\}|\{2,3\})\\
						A(\{2,3\}|\{1,2\}) & A(\{2,3\}|\{1,3\}) & A(\{2,3\}|\{2,3\}) 
\end{bmatrix}.
\] 
Definition~\ref{def:multi} has several implications. First, if~$A$ is square then~$(A^T)^{(k)} = (A^{(k)})^T$, and in particular if~$A$ is symmetric then so is~$A^{(k)}$. 
Also, $A^{(1)}=A$ 
and if~$A \in \mathbb{R}^{n \times n}$ then~$A^{(n)}=\det(A)$.
If~$D$ is an~$n\times n$ diagonal matrix, i.e.~$D=\diag(d_1,\dots,d_n)$ then
$
D^{(k)}=\diag(d_1\dots d_k, d_1\dots  d_{k-1}d_{k+1},\dots,d_{n-k+1}\dots d_n)
$.
In particular, every eigenvalue of~$D^{(k)}$ is the product of~$k$ eigenvalues of~$D$. 
In the special case~$D= p I_n$, with~$p \in \R$, we have that~$(pI_n)^{(k)}=p^k I_r$, with~$r:=\binom{n}{k}$.  
 
The \emph{Cauchy-Binet formula}  (see, e.g.,~\cite[Thm.~1.1.1]{total_book}) asserts   
that 
\be\label{eq:cbf}
(AB)^{(k)}=A^{(k)} B^{(k)}  
\ee
  for any $A \in \R^{n \times p}$, $B \in \R^{p \times m}$, $k \in [1,\min\{n,p,m\} ]$.
  This justifies the term \emph{multiplicative compound}.  
  
When~$n=p=m=k $, Eq.~\eqref{eq:cbf}  becomes the familiar formula~$\det(AB)=\det(A)\det(B)$.
If~$A$ is $n\times n $ and non-singular then~\eqref{eq:cbf}  implies that $ I_n^{(k)}=(AA^{-1})^{(k)}=A^{(k)} (A^{-1})^{(k)}$,
	so~$A^{(k)} $ is also non-singular  with~$(A^{(k)})^{-1}=(A^{-1})^{(k)}   $. 
	Another implication of~\eqref{eq:cbf}
	is that if~$A\in\R^{n\times n}$ with eigenvalues~$\lambda_1,\dots,\lambda_n$
then the eigenvalues of~$A^{(k)}$ are all the~$\binom{n}{k}$ products:
\[
\lambda_{i_1} \lambda_{i_2} \dots \lambda_{i_k}, \text{ with } 1\leq i_1 < i_2 <\dots < i_k\leq n.
\]
	
	The usefulness of the $k$th multiplicative compound in analyzing 
	$k$-contraction follows from the following fact. Fix~$k$ vectors~$x^1,\dots,x^k \in \R^n$. The parallelotope  generated by these vectors (and the zero vertex) is
	\[
	\mathcal{P}(x^1 , \dots,x^k) : =\left \{\sum_{i=1}^k r_i x^i \st r_i \in [0,1] \text { for all } i \right \}.
	\]
	Let
	$
	X:=\begin{bmatrix} x^1&\dots&x^k
	\end{bmatrix} \in \R^{n\times k}.
	$
	The volume of~$\mathcal{P}(x^1,\dots,x^k)$ satisfies~\citep[Chapter~IX]{Gantmacher_vol1}:
	\be\label{eq:voldet}
	\vol (\mathcal{P}(x^1,\dots,x^k))= |X^{(k)} |_2.
	\ee
	Note that since~$X\in\R^{n\times k}$, the dimensions of~$X^{(k)}$ are~$\binom{n}{k}\times 1$, that is, $X^{(k)}$ is a column vector.

	In the special case~$k=n$, Eq.~\eqref{eq:voldet}
	becomes
	\[
	\vol (\mathcal{P}(x^1,\dots,x^n))= |X^{(n)} |_2=|\det(X)|.
	\]
	When the   vertices of the parallelotope
	follow  a linear time-varying dynamics, the evolution of the $k$th multiplicative compound depends on another algebraic construction called the $k$th additive compound. 
\begin{Definition}\label{def:addi_comp}
    The $k$th \emph{additive compound matrix} of a square matrix~$A \in \R^{n \times n}$
is   defined  by
\begin{align} 
    			A^{[k]} := \frac{d}{d \varepsilon}  (I_n+\varepsilon A)^{(k)} |_{\varepsilon=0} =
    			\frac{d}{d \varepsilon}  
    			(\exp(A\varepsilon ))^{(k)}|_{\varepsilon=0}.
\end{align}
\end{Definition}
\begin{Example}
    Suppose that~$A=pI_n$, with~$p\in\R$. Then 
   \begin{align*}
  ( I_n+\varepsilon A)^{(k)}  &=( (1+\varepsilon p  ) I_n)^{(k)} \\
  &=(1+\varepsilon p  )^k I_r, 
   \end{align*}
where~$r:=\binom{n}{k}$, so   
  \begin{align*}   
    	(p I_n)^{[k]} &= \frac{d}{d \varepsilon}  (1+\varepsilon p  )^k I_r |_{\varepsilon=0}\\
    	&=k p I_r.
    	\end{align*} 
\end{Example}

Definition~\ref{def:addi_comp} implies that~$A^{[1]}=A$, 
$A^{[n]}=\tr(A)$,
and that 
\be\label{eq:poyrt}
(I_n+\varepsilon A)^{(k)}= I_r+\varepsilon   A^{[k]} +o(\varepsilon )  .
\ee 
Thus,~$\varepsilon  A^{[k]}$ is the first-order term in the Taylor series of~$(I+\varepsilon A)^{(k)}$. 
Also,   $(A^T)^{ [k]} = (A^{[k]})^T$, and in particular if~$A$ is symmetric then so is~$A^{[k]}$. 

\begin{Example}\label{exa:sdig}
If~$D=\diag(d_1,\dots,d_n)$ then
$
(I+\varepsilon D)^{(k)}=\diag \left ( \prod_{i=1}^k (1+\varepsilon d_i) 
 ,\dots,\prod_{i=n-k+1}^n (1+\varepsilon d_i)    \right  ),
$
  so~\eqref{eq:poyrt} gives 
$
D^{[k]}=\diag( \sum_{i=1}^k d_i 
 ,\dots,  \sum_{i=n-k+1}^n d_i
 )     .
$
In particular, every eigenvalue of~$D^{[k]}$ is the sum of~$k$ eigenvalues of~$D$. 
\end{Example}
 
 More generally,   if~$A\in\R^{n\times n}$ with eigenvalues~$\lambda_1,\dots,\lambda_n$
then the eigenvalues of~$A^{ [ k ] }$ are all the~$\binom{n}{k}$ sums:
\[
\lambda_{i_1} + \lambda_{i_2} + \dots +  \lambda_{i_k} , \text{ with } 1\leq i_1 < i_2 <\dots < i_k\leq n.
\]

It follows from~\eqref{eq:poyrt} and the properties of the multiplicative compound that~$
(A+B)^{[k]}= A^{[k]}+B^{[k]} 
$ for any~$A,B\in\R^{n\times n}$,  
thus justifying the term \emph{additive compound}. In fact, the mapping~$A\to A^{[k] }$ is linear~\citep{schwarz1970}.

Below we  will use the following  relations. Let~$A \in \R^{n \times n}$, and~$k\in[1,n]$. If $T \in \R^{n \times n}$ is invertible, then
\begin{equation}\label{eq:k_mul_coor_trans}
(TAT^{-1})^{ (k) } = 	T^{(k)}A^{ (k) } (T ^{(k)}  ) ^{-1}   , 
\end{equation}
and combining this with Definition~\ref{def:addi_comp} gives
\begin{equation}\label{eq:k_add_coor_trans}
(TAT^{-1})^{[k]} = 	T^{(k)}A^{[k]} (T ^{(k)}  ) ^{-1}   
\end{equation}
(see, e.g.,~\cite{comp_long_tutorial}). 

  For more on the applications of compound matrices  to
systems and control theory, see e.g.~\citep{wu2021diagonal, margaliot2019revisiting, ofir2021sufficient, ron_DAE,grussler2022variation,LI1999191}, and the recent tutorial by~\cite{comp_long_tutorial}.

\subsection{Matrix measures}
A  norm~$|\cdot|:\R^n\to\R_+$
 induces a  matrix norm~$\|\cdot\|:\R^{n\times n}\to\R_+$   
defined by~$\|A\| := \max_{|x|=1} |Ax| $, and a   matrix measure   $\mu(\cdot):\R^{n\times n}\to\R $    
defined by~$
\mu(A) := \lim_{\varepsilon \downarrow 0} \frac{\|I + \varepsilon A\| - 1}{\varepsilon} .
$
The matrix measure is sub-additive, i.e.~$\mu(A+B)\leq\mu(A)+\mu(B)$, and~$\mu(c I_n)=c$ for any~$c\in \R$. 

Denote the~$L_2$  norm  by
$|x|_2:=(x^Tx)^{1/2}$.
The corresponding matrix norm
is~$\| A \|_2=\left (\lambda_{\max}(A^TA)\right)^{1/2} $, 
and the 
corresponding matrix 
measure is~\citep{vidyasagar2002nonlinear}:
\begin{equation}\label{eq:matirxm} \begin{aligned} 
\mu_2(A) = (1/2) \lambda_{\max}\left(  A + A^T \right) ,   
\end{aligned} \end{equation}
where~$\lambda_{\max} (S)$ denotes the  largest eigenvalue of the symmetric matrix~$S$.

For an invertible matrix~$H \in\R^{n\times n}$ the scaled~$L_2$ norm is defined by~$|x|_{2,H} :=|H x|_2$, and the induced matrix measure is 
\begin{align}\label{eq:weight_mat_meas}
    \mu_{2,H}(A)& =   
    \mu_2(H A H^{-1} )\nonumber\\& 
    =    (1/2)   \lambda_{\max}\left(H A  H^{-1}  + 
    (H A H^{-1} )^T  \right).
\end{align}

As shown in~\citep{kordercont}, a sufficient condition for the system~$\dot x=f(t,x)$ to be $k$-contractive  is that~$\mu((J(t,x))^{[k]})\leq-\eta<0$ for all~$t,x$, where~$J:=\frac{\partial}{\partial x}f$. For~$k=1$, this reduces to the standard sufficient condition for contraction, namely,  $\mu(J (t,x)) \leq -\eta<0$ for all~$t,x$.
Note that if~$A,T\in\R^{n\times n}$, with~$T$ non-singular, then
\begin{align}\label{eq:mu2add}
\mu_{2,T^{(k)}  } (A^{[k]} ) & = \mu_{2  } (T^{(k)}A^{[k]} (T^{(k)}) ^{-1} )\nonumber\\
&=\mu_2(  (TAT^{-1})^ { [k]}    ). 
\end{align}

\section{Main results}\label{sec:main}
In this section, we derive a sufficient condition for $k$-contraction of the closed-loop system~\eqref{eq:closed_loop1}.
We assume that the nonlinearity~$\Phi$ is continuously differentiable and denote its Jacobian
by~$J_\Phi(t,y) := \frac{\partial \Phi}{\partial y}(t,y)$. The Jacobian of~\eqref{eq:closed_loop1} is then
\begin{equation}\label{eq:closed_loop_jac}
    J(t,x): = A - BJ_\Phi(t,Cx)C.
\end{equation}

We can now state our main result. 
For a symmetric matrix~$S\in\R^{n\times n}$,  we denote its ordered eigenvalues as~$\lambda_1(S)\geq\dots\geq \lambda_n(S)$.
\begin{Theorem}\label{thm:lure_suff}
Consider the Lurie system~\eqref{eq:closed_loop1}.
	Fix~$k \in [1,n]$.
	Suppose  that
	there exist~$\eta_1,\eta_2\in \R$ and~$P \in \R^{n \times n}$, where~$P =QQ $ with~$Q \succ 0$, such that
	\begin{align}\label{eq:k_riccati_Pk}
    &    P^{(k)}A^{[k]}  + (A^{[k]})^TP^{(k)} + \eta_1 P^{(k)} \\
        &+ Q^{(k)}\left((QBB^TQ)^{[k]} + (Q^{-1}C^TCQ^{-1})^{[k]}\right)Q^{(k)} \preceq 0 , \nonumber
    \end{align}
and
	\begin{equation}\label{cond:k_smallgain}
		\sum_{i=1}^k \lambda_i\left( Q^{-1}C^T \left (
  (J_\Phi^T(t,y)  J_\Phi(t,y) - I_q
\right   )  CQ^{-1} \right ) \leq - \eta_2 ,
	\end{equation}
	for all $t \ge 0, y \in \R^q$. Then the Jacobian of the closed-loop system~\eqref{eq:closed_loop1} satisfies
	\[
	\mu_{2,Q^{(k)}}( J^{[k]}(t,x)) \leq  -(\eta_1 + \eta_2)/2 \text{ for all } t \ge 0, x \in \R^n.
	\]
  In particular, if~$\eta_1 + \eta_2 > 0$, then the closed-loop system~\eqref{eq:closed_loop1} is $k$-contractive with rate~$(\eta_1 + \eta_2)/2$ w.r.t. the scaled $L_2$ norm~$|z|_{2,Q^{(k)}}=|Q^{(k)}z|_2$.
\end{Theorem}

 \begin{Remark}
Note that when~$k=1$, Eq.~\eqref{eq:k_riccati_Pk} holds for some~$\eta_1 > 0$ if and only if the familiar ARI
\be\label{eq:fari}
 P A  + A^TP  + P BB^T P  +  C^TC  \prec 0
\ee
holds.  Assuming the LTI subsystem is minimal,~\eqref{eq:fari}  holds   if and only if $A$ is Hurwitz and the $H_\infty$ norm of the LTI subsystem is less than~1. Similarly,~\eqref{cond:k_smallgain} holds for any~$\eta_2 > 0$ if~$\|J_\Phi\|_2 \le 1$, so in this case  Thm.~\ref{thm:lure_suff} becomes a small-gain sufficient condition for standard contraction.
\end{Remark}

 The rest of this section is devoted to the proof
of Thm.~\ref{thm:lure_suff}. We require the following   result.
\begin{lem}\label{lem:MN_matrices}
Fix~$M \in \R^{n \times m}$,
$N \in \R^{m \times n}$, and~$k\in\{1,\dots,n\}$.
Then
\[
( - M N - N^T M^T  -  N^T N )^{[k]}\preceq (M M^T) ^{[k]}. 
\]
\end{lem}
\begin{pf}
    The  identity
    \[
    M N + N^TM^T = (M^T + N)^T(M^T + N) - MM^T - N^T N 
    \]
    gives
    \[
   Z:=- M M^T  - M N - N^T M^T  -  N^T  N \preceq 0.
   \]
    Thus,  $Z$ is symmetric with all (real)
    eigenvalues smaller or equal to zero. Hence, the same properties hold   
      for~$Z^{[k]}$, so
    \[
   Z^{ [ k]}= \left (-  M M^T  - M N - N^T M^T  -  N^T  N \right  )^{[k]}\preceq 0 , 
   \]
   and this completes the proof. 
\end{pf}

We can now prove Thm.~\ref{thm:lure_suff}.
\begin{pf}
	Let $R := Q J  Q^{-1} + Q^{-1} J  ^TQ$, with~$J$ defined  in~\eqref{eq:closed_loop_jac}. Then
    \begin{align*}
        R^{[k]} &=\left (Q(A - BJ_\phi C)Q^{-1} + Q^{-1}(A - BJ_\phi C)^TQ \right )^{[k]} \\
        &=
        \left (QAQ^{-1} + Q^{-1}A^T Q \right )^{[k]}\\&
        -\left (QBJ_\phi C Q^{-1} + Q^{-1}  C^TJ_\phi^T B^T Q ) \right )^{[k]}.
    \end{align*}
    Multiplying~\eqref{eq:k_riccati_Pk} on the left- and on the right-hand side  by~$(Q^{(k)})^{-1}$, and using~\eqref{eq:k_add_coor_trans} gives
    \begin{multline}\label{eq:kARIcond_Q}
        (QAQ^{-1} + Q^{-1}A^T Q)^{[k]} \preceq \\ -\eta_1 I_r - (QBB^TQ + Q^{-1}C^TCQ^{-1})^{[k]}, 
    \end{multline}
    so
    \begin{align}\label{eq:rsofar}
        R^{[k]} &\preceq  -\eta_1 I_r -\left (QBB^TQ + Q^{-1}C^TCQ^{-1  }\right)^{[k]} \nonumber \\
        & -\left (QBJ_\phi C Q^{-1} + Q^{-1}  C^TJ_\phi^T B^T Q   \right )  ^{[k]}.
    \end{align}
    It follows from Lemma~\ref{lem:MN_matrices} with~$M=Q B  $
    and~$N=J_\phi CQ^{-1}$ that
    \begin{align*}
        ( - Q B J_\phi  C Q^{-1}  &-  Q^{-1} C^T 
        J_\phi^T B^T Q   -  Q^{-1} C^T J_\phi^T J_\phi  CQ^{-1}  )^{[k] } \\
        &\preceq ( Q B  B^T Q   ) ^{[k]} ,  
    \end{align*}
    and combining this with~\eqref{eq:rsofar} gives
    \begin{align}\label{eq:mulj}
        R^{[k]} &\preceq  -\eta_1 I_r 
        + \left (Q^{-1}C^T  (J_\phi ^T J_\phi  -I_q) C Q^{-1}  
        \right)^{[k]} .
    \end{align}
Using~\eqref{cond:k_smallgain}, we get
    \begin{equation*}
       \mu_2(R^{[k]}) 
        \le -\eta_1 - \eta_2,
    \end{equation*}
and  the definition of~$R$  
gives~$ 2\mu_{2,Q^{(k)}}(J^{[k]}) = \mu_2(R^{[k]}) 
        \le -\eta_1 - \eta_2.
$
In particular, if~$\eta_1 + \eta_2 > 0$ then    the closed-loop system is~$k$-contractive with rate~$ (\eta_1 + \eta_2)/2$ w.r.t. the scaled~$L_2$ norm~$|z|_{2,Q^{(k)}}= |Q^{(k)} z|_2$. This completes the proof of Thm.~\ref{thm:lure_suff}. \qed 
 \end{pf}
 
 \begin{Remark}
Define the symmetric matrix
\begin{align*}S &:= QAQ^{-1} +Q^{-1}A^TQ   + {\eta_1}{k^{-1}} I_n +  QB B^TQ \\&+  Q^{-1}C^TCQ^{-1} .
\end{align*}
Then condition~\eqref{eq:k_riccati_Pk} can be written more succinctly as~$S^{[k]}\preceq 0$, that is, $\sum_{i=1}^k\lambda_i (S)\leq 0$.
\end{Remark}

\begin{Remark}\label{rem:scalar_P_C}
    Note that for the particular case~$P=pI_n$, with~$p>0$, and~$C=I_n$ (i.e., the LTI output is~$y=x$), we have~$Q=p^{1/2} I$, and Eq.~\eqref{eq:mulj} gives
 	\begin{align*} 
		2\mu_{2,Q^{(k)}}(J^{[k]})  
			\le & 
			  -\eta_1 - p^{-1} k + p^{-1}\sum_{i=1}^k \sigma_i^2(J_\Phi) . 
	\end{align*}
Thus, in this  case 
a sufficient condition for $k$-contraction is  
\be\label{eq:sumsuff}
\sum_{i=1}^k \sigma_i^2(J_\Phi (t,y)  )  < k+ {\eta_1}{p } \text{ for all } t\geq 0,y\in\R^n.
\ee
 \end{Remark}
 
\section{An Application: $k$-contraction in a networked system}
Consider the nonlinear  networked dynamical system
\begin{equation}\label{eq:net_sys}
    \dot x = -\alpha x + Wf(x),
\end{equation}
where $x \in \Omega \subseteq \R^n, \alpha > 0, W \in \R^{n \times n} $ is a matrix of interconnection weights, and $f : \R^n \to \R^n$. In the context of neural network models, the~$f_i$s are   the  neuron  activation functions. We assume that the state space~$\Omega$ is convex and that~$f$ is continuously differentiable. Let~$J_f (z) := \frac{\partial f}{\partial z}(z)$. 

Intuitively speaking, it is clear that  a larger~$\alpha$ makes the system ``more stable''.
The next result formalizes this by providing
a sufficient condition for $k$-contraction.

\begin{cor}\label{cor:net_k_contract}
  Consider~\eqref{eq:net_sys} with~$\alpha>0$.   Fix~$k \in [1,n]$. If 
    \begin{equation}\label{cond:net_k_smallgain}
    \|J_f(x)\|_2^2 \sum_{i=1}^k \sigma_i^2(W) < \alpha^2 k  \text{ for all } x \in \Omega,
    \end{equation}
    then~\eqref{eq:net_sys} is $k$-contractive. 
    Furthermore, if~$f$ is uniformly bounded and~\eqref{cond:net_k_smallgain} holds with $k=2$  then every trajectory of~\eqref{eq:net_sys} converges to an equilibrium point (which is not necessarily unique).
\end{cor}
\begin{Remark}
Note that if~\eqref{cond:net_k_smallgain} holds for some~$k\in [1,n]$ then it holds for any 
$\ell\geq k$. Thus, if the sufficient condition for~$k$-contraction holds then the system is also~$\ell$-contractive for any~$\ell\geq k$. This agrees with the results in~\cite{kordercont,wu2020generalization}. Note also that if~$W=0$ or~$f(x)=0$ then~\eqref{cond:net_k_smallgain} holds for~$k=1$ (and thus for any~$k\in [1,n]$). This is reasonable,  as in this case we have~$\dot x=-\alpha x$, and this is indeed $k$-contractive for any~$k\geq 1$. 
\end{Remark}
\begin{pf}
The proof is based on Thm.~\ref{thm:lure_suff}. We first represent the networked system~\eqref{eq:net_sys} as a Lurie system. By~\eqref{cond:net_k_smallgain}, there exists~$\gamma$ such that
\be\label{eq:gammprop}
 0<\gamma < \alpha \text{ and } \|J_f(y)\|_2^2 \sum_{i=1}^k \sigma_i^2(W) < \gamma^2 k.
\ee
 We can represent~\eqref{eq:net_sys} as 
 the interconnection of the LTI system with $(A,B,C)=(-\alpha I_n, \gamma I_n, I_n)$ and the nonlinearity~$\Phi(y) := -\gamma^{-1}Wf(y)$, that is,
\begin{align}\label{eq:lui}
\dot x&= -\alpha x + \gamma u,\nonumber\\
u&= \gamma^{-1} Wf(x ).
\end{align}
For this Lurie system, there exist~$Q \succ 0$ with~$P=QQ$ and~$\eta_1>0$ such that~\eqref{eq:k_riccati_Pk} holds if and only if
\begin{equation}\label{eq:k_riccati_net}
    -2k\alpha P^{(k)} + Q^{(k)}(\gamma^2 P + P^{-1})^{[k]}Q^{(k)} \prec  0.
\end{equation}
Taking~$P = p I_n$, with~$p>0$, Eq.~\eqref{eq:k_riccati_net} simplifies to
\begin{equation}\label{eq:k_riccati_scalar}
    (-2\alpha + \gamma^2 p + p^{-1}) k p^k < 0,
\end{equation}
which indeed admits a solution~$p>0$ since~$\alpha > 0$ and~$\gamma < \alpha$. We conclude that there exists a matrix~$P = p I_n$, with~$p>0$, and a scalar~$\eta_1 > 0$ for which~\eqref{eq:k_riccati_Pk} holds.

We now show that~\eqref{cond:net_k_smallgain} implies that~\eqref{cond:k_smallgain} holds for some~$\eta_2>0$.
Since~$P=pI_n$ and~$C=I_n$, we may apply the result in Remark~\ref{rem:scalar_P_C}. Recall that for any~$A\in\R^{m\times p},B\in\R^{p\times n}$, we have
\be\label{eq:sigmai2}
    \sum_{i=1}^k \sigma_i^s(AB) \leq
    \sum_{i=1}^k (\sigma_i(A)\sigma_i(B))^s  
\ee
for any $k\in[1,\min\{m,p,n\}]$, $s>0$
\citep[Thm.~3.3.14]{Horn1991TopicsMatrixAna}. Consider
\begin{align*}
    \sum_{i=1}^k \sigma_i^2(J_\Phi)
& = \sum_{i=1}^k \sigma_i^2 (-\gamma^{-1} W J_f)\\
&\leq\gamma^{-2}  \sum_{i=1}^k \sigma_i^2 (  W  )\sigma_i^2 (    J_f)\\
&<k,
\end{align*}
where the first  inequality follows from~\eqref{eq:sigmai2}, 
and the second   from~\eqref{eq:gammprop}. 
We conclude that the sufficient condition~\eqref{eq:sumsuff} holds,
and  Thm.~\ref{thm:lure_suff} implies that~\eqref{eq:net_sys} is $k$-contractive.

Suppose now that~\eqref{cond:net_k_smallgain} holds with~$k=2$. Then~\eqref{eq:net_sys} is 2-contractive. If in addition~$f$ is uniformly bounded, then all the  trajectories of~\eqref{eq:net_sys} are bounded, and by known  results on time-invariant  2-contractive systems~\citep{li1995} we then have that all trajectories   converge to an equilibrium point. This completes the proof of Corollary~\ref{cor:net_k_contract}. \hfill{\qed}
\end{pf}

A particular example of a   system in the form~\eqref{eq:net_sys} is the famous Hopfield network~\citep{hopfield_net}. This network has been used as an  associative memory, where each equilibrium corresponds to a stored pattern (see, e.g.,~\cite{krotov2016}). Hence, the network is    multi-stable and thus not contractive  w.r.t. any norm. 
  Corollary~\ref{cor:net_k_contract} may   be used to prove that a
  Hopfield network is~$k$-contractive for~$k>1$. To show this,  consider~\eqref{eq:net_sys} with~$n=10, \alpha = 1/2, W = 1_n 1_n^T$, where~$1_n \in \R^n$ is a column vector of ones, and~$f(x) = 0.07 \tanh(x)$. In this case,~\eqref{eq:net_sys} has at least 3 equilibrium points: $e^1=0$,     $e^2\approx   1.1403 \cdot 1_n$, and~$e^3=-e^2$, so it is certainly not $1$-contractive. However, condition~\eqref{cond:net_k_smallgain} holds for~$k=2$   since~$\| J_f(x)\|_2^2 \leq 0.07^2$,
  $\sum_{i=1}^2\sigma_i^2(W) = 100$, and~$\alpha^2 k =1/2$. 
Thus,   the system is 2-contractive. Since~$f$ is also uniformly bounded, we conclude that  all trajectories converge to an equilibrium point. Note that when using the network as an associative memory, such a property is very useful. 

\section{Conclusion}
We derived a sufficient condition for~$k$-contraction of Lurie systems. For~$k=1$, this reduces to the standard sufficient condition for contraction. However, often Lurie systems admit more than a single  equilibrium point, and are thus not contractive (that  is, not~$1$-contractive) w.r.t. any norm. 
Our condition may still be used to guarantee a well-ordered behaviour of the closed-loop system. For example in the time-invariant case, establishing that the system is~$2$-contractive implies that 
any bounded solution converges  to an equilibrium, that is not necessarily unique. Such a property is important, for example, in dynamical models of associative memories, where every equilibrium corresponds to a stored memory.

\subsection*{Acknowledgments} 
We thank the anonymous reviewers for their helpful comments.


\end{document}